# Solid-state qubits in moiré superlattices

Zhigang Song[1*], Péter Udvarhelyi[2], Yidan Wang[3], Prineha Narang[4,5*]

1 John A. Paulson School of Engineering and Applied Sciences, Harvard University, Cambridge, Massachusetts 02138, United States

2 Department of Chemistry and Biochemistry, University of California, Los Angeles, Los Angeles, CA, USA

3 Physics Department, Harvard University, Cambridge, Massachusetts 02138, USA

4 Division of Physical Sciences, College of Letters and Science, University of California, Los Angeles, Los Angeles, CA, USA

5 Department of Electrical and Computer Engineering, University of California, Los Angeles, Los Angeles, CA, USA

Qubits are the fundamental units in quantum computing, but they are also pivotal for advancements in quantum communication and sensing. Currently, there are a variety of platforms for qubits, including cold atoms, superconducting circuits, point defects, and semiconductor quantum dots. In these systems, each qubit requires individual preparation, making identical replication a challenging task. Constructing and maintaining stable, scalable qubits remains a formidable challenge, especially for solid-state qubits. The race to identify the best one remains inconclusive, making the search for new qubits a welcome endeavor. Our study introduces moiré superlattices of twisted bilayer materials as a promising platform for qubits due to their tunability, natural patterns, and extensive materials library. Our first-principles calculations reveal that when the twist angle between the two layers is sufficiently small, these materials foster identical, localized quantum wells within the moiré superlattices. Each quantum well accommodates a few dispersionless bands and localized states, akin to the discrete energy levels of an alkali atom. Existing experimental techniques allow for individual initialization, manipulation, and readout of the local quantum states. The vast array of 2D materials provides a multitude of potential candidates for qubit exploration in such systems. Due to their inherent scalability and uniformity, our proposed qubits present significant advantages over conventional solid-state qubit systems.

Quantum computers have the potential to revolutionize various industries, solving complex problems that are currently beyond the capabilities of classical computers.[1] Interest in quantum computation has surged recently, driven by demonstrations of quantum supremacy and advancements in error correction techniques.[2-4] Quantum bits (qubits) are the fundamental units of quantum computers, where information is processed through quantum gate operations. Beyond quantum computing, qubits also play crucial roles in quantum sensing, precision measurements, and quantum communication.[5] Over the past decades, qubits have been implemented in various platforms, such as superconducting circuits,[6] trapped ions,[7] neutral atoms,[8] molecular qubits,[9] semiconductor quantum dots[10], and semiconductor defect centers[11]. Among the different types, qubits in solid-state materials have garnered special attention. As solid-state qubits are naturally embedded in their host crystals, they do not require ultrahigh vacuum, have less stringent temperature requirements than atomic qubits, and are compatible with existing semiconductor technologies. However, they often require individual preparation, making identical replication challenging. It is also difficult to scale up their number in a single computation unit. Achieving tunable, scalable, and identical solid-state qubits remains a formidable task. To overcome these challenges, significant efforts are being made to explore new qubit designs based on solid-state materials, and the race to develop viable quantum technologies continues in both experiments and theory.[12-16]

In the field of condensed matter physics, twisted bilayer materials (or moiré superlattices) have recently emerged as a fascinating platform.[17] By tuning the twist angle and applying external fields, scientists explored various exotic quantum phenomena.[18,19] Twisted bilayer materials are considered promising quantum simulators for the Hubbard model with narrow bands (bandwidth ~100 meV).[20-23] Previous research indicated the existence of ultra-narrow bands (bandwidth ~10 meV) under small twist angles.[24-26] If the twist angle is further reduced, the kinetic energy in most twisted bilayer materials is believed to vanish, resulting in dispersionless flat bands. Due to methodological limitations, there has been little exploration of bands with widths below 1 meV. First-principles calculations of these dispersionless bands in twisted bilayer materials are challenging due to the large number of atoms in the periodic supercell (~10,000).[27] By developing specific density functional theory (DFT) methods for large-scale material systems, our recent work has overcome these methodological challenges and indeed found dispersionless bands in moiré superlattices.[28,29] Dispersionless bands consist of atom-like localized states instead of propagating states.

In our current work, we study the quantum properties of dispersionless bands and operations of these states, exploring their potential as spin-based qubits for universal quantum computation. In our DFT calculations, moiré superlattices have dispersionless bands with narrow linewidth and longer lifetimes. The local states in the moiré superlattices are self-organized and exhibit excellent identicality for each site. Due to the periodicity, freedom of twist angle, and extensive materials library, moiré superlattices naturally enable identical, scalable, and highly tunable qubits. Additionally, the localized states introduced into the bandgap of the host material make the moiré spin qubit optically active with relatively strong emission. This allows for optical spin initialization and readout at elevated temperatures. Moiré superlattices provide an atomically thin solid-state platform for quantum information sciences and scalable integrated spin-photonic quantum systems.

**Electronic structure calculations of twisted bilayer materials**

To describe the principles of qubit operation in moiré superlattices, we use density functional theory (DFT) to investigate their electronic structure in the example formed by twisting two PbS layers, as illustrated in Fig. 1a. After relaxation due to the release of atomic forces, the AA stacking (where atoms stack directly on top of the same atomic species) zone shrinks to form a series of self-organized quantum dots, while the AB stacking (where atoms stack on top of different atomic species) zone expands. An example with a twist angle of 2.66° is depicted in supplementary Fig. S1. As the twist angle ($\vartheta$) decreases, the inter-site distance increases as $R = \frac{a}{\sqrt{2}\sin(\vartheta/2)}$, where $a$=4.17 Å is the lattice constant of the PbS monolayer. The quantum dots maintain an almost constant radius of around 1 nm for different twist angles. Our DFT results are in good accordance with previous experiments.[30]

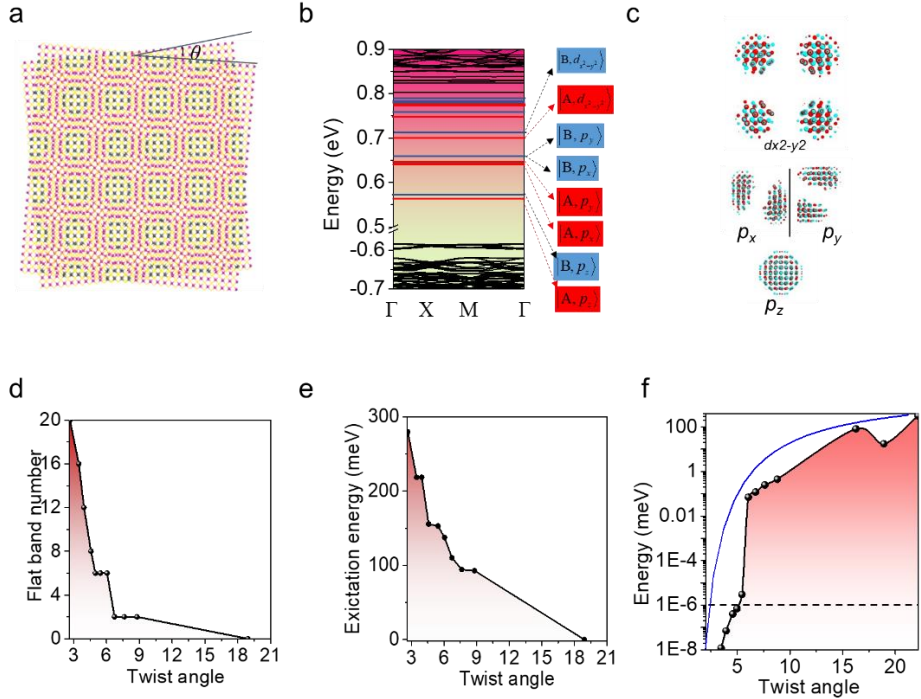

**Figure 1. Electronic structure of twisted bilayer PbS. a** Illustration of twisted bilayer PbS. **b** Calculated band structure under a twist angle of 2.66° (left). Red and blue colors indicate the states located at A and B sublattices, respectively. **c** Calculated wave functions of dispersionless bands near the Fermi energy, corresponding to the states colored red in panel b. Color represents the phase of the wavefunction. **d** Number of localized states as a function of the twist angle. **e** The energy separation between the lowest dispersed band and lowest dispersionless band ($\Delta_b$) as a function of the twist angle. **f** Black curve represents the bandwidth of the first unoccupied band as a function of the twist angle. Blue curve shows the analytical exponential function. Dashed line indicates the bandwidth limit that can be considered as zero dispersion.

The energy bands (energy dispersion as a function of momentum) become increasingly flat as the twist angle decreases as shown in Extended Data Fig. 1. At a twist angle of 4.58°, the calculated width of the flat band above the Fermi level is 0.0007 meV, which is negligible given the possible error in DFT calculations. This band can be considered dispersionless. The bandwidth decreases rapidly with the twist angle. The calculated band structure with a twist angle of 2.66° is shown in Fig. 1b. The dispersionless bands have atom-like localized wavefunctions and exhibit specific symmetry (see Fig. 1c). Above the Fermi level, the wavefunctions of the flat bands are characterized by $p_z$, $p_x$, $p_y$, $d_{x2-y2}$, and other states. $p_x$ and $p_y$ states are degenerate, forming $|l_z = \pm\rangle = |p_x\rangle \pm i|p_y\rangle$. The energy separation between $p_z$ and $p_x$ ($p_y$) states is 78 meV. The energy separation between the $p_z$-state and the lowest dispersed band (with a bandwidth larger than 1 meV) is as large as $\Delta_b$=277 meV. This suggests that the tunneling probability $e^{-\frac{\Delta_b}{k_B T}}$ to escape from a quantum dot at low temperatures ($T$) is very small. Twisted bilayer PbS adheres to the $D_4$ point group symmetry. As shown in Extended Data Table 1. The $p_z$ state belongs to the $A_2$ irreducible representation of the $D_4$ point group, while the $p_x$ and $p_y$ states belong to the $E$ irreducible representation. Symmetry considerations and angular momentum conservation will lead to a selection rule for dipole excitation between two localized states.

In addition to twisted bilayer PbS, our DFT calculations in Extended Data Fig. 2 show that other materials, such as twisted bilayer $In_2Se_3$ and $Bi_2Se_3$, also exhibit dispersionless bands, but the localized states are below the Fermi level. Both twisted $In_2Se_3$ and $Bi_2Se_3$ have a triangular lattice of quantum dots. For $In_2Se_3$, when the twist angle is smaller than 3.14°, the bandwidth declines below 1.0 meV. In twisted bilayer $Bi_2Se_3$, the bandwidth is smaller than 0.5 meV, when the twist angle is 2.65°. Besides this work, ultra-flat bands rather than dipersionless bands have been reported in twisted bilayer h-BN, and twisted bilayer $MoS_2$.[31,32] In experiments, the twist angle can be as small as 0.08° or even smaller.[33,34] However, the quantum applications and operations of dispersionless bands have rarely been explored.

**General theory**

The conclusions of our numerical calculations can be generalized to any other materials. Dispersionless bands and atom-like localized states are generally present in twisted heterostructures or homostructures, if the twist angle is sufficiently small, except in special cases of linear dispersion. After twisting, the relaxation process simultaneously results in a periodic displacement vortex in any materials (illustrated in Supplementary Fig. 2a). The strain distribution and periodic pattern of quantum dots under different symmetries are discussed in Supplementary Note 1 and Supplementary Figs. 2&3.

The electronic structure of a layer in twisted bilayer materials can be described by a model of a particle inside a symmetric quantum well as follows:

$$H = -\frac{\hbar^2 \nabla^2}{2M_{eff}} + V(\vec{r}), \qquad (1)$$

where $M_{eff}$ is the effective mass at the band edge of the original layers, $\hbar$ is the reduced Planck constant and the potential $V(\vec{r})$ is determined by the interlayer coupling and atomic relaxation after twisting (see the method part). In twisted bilayer PbS quantum-well depth is estimated as 201 meV and 167 meV for conduction and valence bands, respectively. The problems of bound states in square, hexagonal[35], and triangle[36] quantum wells in Eq.1 have been well addressed in previous works. Detailed information on square lattices can be found in Supplementary Note 2 for reference. The dispersionless bands correspond to the bound states in a quantum well. If the potential depth and effective mass are sufficiently large, dispersionless bands (or bound states) will form at the sites of the moiré superlattices no matter what kind of quantum wells.

Based on the symmetry, we conducted a data-mining search for suitable layered compounds from the materials database of more than 5000 layered materials.[37] The materials database is extensive and offers a promising avenue for future refinement of qubits. The original layered materials and their possible moiré lattices according to symmetries are listed at the end of Supplementary Note 1. Dispersionless bands should reside within the bulk bandgap to avoid coupling with dispersed bands, and thus we filter out two-dimensional semiconductors with band gaps ranging from 0 to 5 eV. The well depth and vortex distortion are correlated with the interlayer binding energy as shown in extended data Fig. 3. In the high throughput calculations, interlayer binding energy is given to evaluate the materials. Generally, the binding energy in vdW (non-vdW) materials is typically less (much larger) than 25 meV/Å². Non-vdW materials, for example twisted bilayer PbS, tend to have strong vortex distortions and deep potential wells, with large energy separations between different flat bands. However, atomic defects and unexpected strains in twisted bilayer vdW materials are usually smaller in experiments.

The underlying physics of band dispersion can be captured by a tight-binding model based on periodic quantum dots with a certain symmetry group. We assume there is a potential well localized at each quantum dot labeled by $m$. Each potential well can support multiple states with different orbitals and spins. The Hamiltonian is as follows (illustrated in Fig. 2a):

$$H_0 = \sum_{m\alpha} \varepsilon_m c^\dagger_{m\alpha} c_{m\alpha} + \sum_{m \neq n\alpha} t_{mn} c^\dagger_{m\alpha} c_{\alpha n} + \sum_{mn\alpha\beta} J_{mn} c^\dagger_{\alpha m} c_{\alpha m} c^\dagger_{\beta n} c_{\beta n} + \sum_{mn\alpha\beta} J'_{mn} c^\dagger_{\alpha m} c_{\alpha n} c^\dagger_{\beta m} c_{\beta n} , \qquad (2)$$

where states inside a potential well are labeled by $\alpha$ or $\beta$, $\varepsilon_m$ is the onsite energy, $t_{mn}$ is the hopping integral, $c^\dagger_{\alpha m}$ and $c_{\alpha m}$ are the creation and annihilation operators for the quantum state $\alpha$ on the $m_{th}$ site, respectively. The first and second terms are the single-electron interaction. The third term is the Coulomb interaction component of two-electron interaction, while the last term is the exchange interaction. We temporally include the last two terms in the method of mean-field theory. The parameters in the tight-binding model are not independent according to the symmetry of the system. As shown in Fig. 2b, the band structure near the Fermi level can be well captured by the tight-binding model.

The kinetic energy is related to the $t_{mn}$ hopping term, which is determined by the overlap between the bound states on different sites. Hopping usually decays exponentially with the distance as $t_{mn} \sim F(R)e^{-\chi R}$ according to the overlap integration theory.[38] Thus, the dispersion near the band structure fast decays to zero as the twist angle decreases. The DFT calculated bandwidth of the $p_z$-state and its comparison with the hopping parameter $t_{mn} \sim F(\frac{a}{\sqrt{2}\sin(\vartheta/2)})e^{-\alpha \frac{a}{\sqrt{2}\sin(\vartheta/2)}}$ is shown by the black and blue curves in Fig. 1f, respectively.

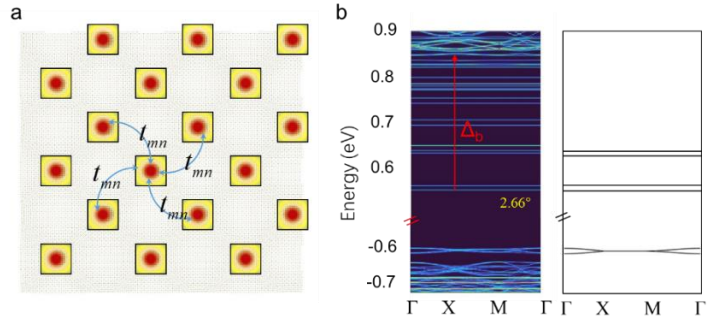

**Figure 2. Quantum-well model and the tight-binding model. a** Quantum-well states and electron hopping. **b** Comparison of DFT band structure (left) and band structure calculated by the tight-binding model with $t$=32.5 meV. $\Delta_b$ represents the scattering barrier.

We included electron-electron interaction in the last two terms of Eq. 2 using the mean-field theory method. Exactly solving the ground states of materials with the many-body description of the electron-electron interaction remains a long-standing challenge in the field of condensed matter physics. In the special case of zero dispersion, the many-body ground states can be solved. The electron-electron interaction does not delocalize the wavefunctions near the Fermi levels, as detailed in Supplementary Note 3. The electron-electron interaction may induce an energy shift, but it does not affect the qubit formation hypothesis discussed in the following section.

**Qubit manipulation**

To use the localized states as qubits in quantum computing, we need to construct protocols for their initialization, coherent manipulation, and readout. As shown in Fig. 3a, qubits are built on a few dispersionless bands that lie deep in the bulk band gap, so their interaction with the dispersed states is suppressed. When a small magnetic field ($B$) is applied (technically, we can use magnets, such as $Nd_2Fe_{14}B$ or $SmCo_5$), a dispersionless band will split into two spin-polarized energy levels, $\Delta_B = -g\mu_B B \sim -14 GHz/T$. $g$ is the Landé $g$-factor, $\mu_B$ is the Bohr magneton. This two-level system $(|\uparrow\rangle, |\downarrow\rangle)^T$ can serve as the computational subspace of a single qubit as shown in Fig. 3b. After an electron injection, the system is initialized in its negative charge state and spin-half spin state. In the regime of dispersionless bands, the tunneling between two neighboring quantum dots vanishes. Any potential tunneling path is bridged by the dispersed bands, as illustrated in Extended Data Fig. 4. The calculated energy barrier between two adjacent artificial atoms is 278 meV in twisted bilayer PbS with a twist angle of 2.66°. As the twist angle decreases, the barrier depth ($\Delta_b$) increases further (see Fig. 1d). Electron tunneling between these sites is exceedingly difficult once prepared, ensuring the stability of the qubit state. The coherence time and stability can be comparable to those of Nitrogen-Vacancy (N-V) centers in diamond, given the similarity in their material structures.

We propose to use electrostatic operation to initialize the qubits. The exact number of electrons in a localized state can be controlled using an electrostatic gate, which consists of an insulating dielectric layer (such as h-BN or $Al_2O_3$) and a metal layer as shown in supplementary Fig. 5. The target local state was biased to control the energy to make sure $\varepsilon_\uparrow > E_f > \varepsilon_\downarrow$. At a low temperature, the state $|\downarrow\rangle$ below the Fermi level is occupied by the tunneling current. The tunneling between $|\uparrow\rangle$ and the reservoir is not allowed. This technique of electrostatic gate has been widely tested in semiconductor-quantum-dot spin qubits.[39] The efficiency of an electrostatic gate based on $Al_2O_3$ on 2.66°-twisted PbS is estimated using DFT calculations, as shown in Extended Data Fig. 5. The energy shift decreases linearly with the applied external voltage. The inter-dot distance, for example 12.75 nm in 2.66° twisted PbS, is sufficiently large to enable the initialization of individual quantum dots using a tailored electrostatic gate design.

To perform single-qubit operations The Hamiltonian of spin-polarized band is written as $H_o = |\uparrow\rangle \varepsilon_\uparrow \langle\uparrow| + |\downarrow\rangle \varepsilon_\downarrow \langle\downarrow|$ with $\Delta_B = \varepsilon_\uparrow - \varepsilon_\downarrow$. A local GHz microwave field can be used to coherently manipulate the electron spin for performing quantum gates. Each qubit typically exhibits a splitting of several hundred MHz to a few GHz, allowing a complete set of gate operations for a single qubit.

The access to several individually addressable orbital states in the moiré systems allows for optical initialization, readout, and detection. The proposed protocol is shown in Fig.3c. The protocol relies on the optically allowed transition between the $p_z$ and $p_x/p_y$ orbital states, similar to all-optical spin initialization methods already demonstrated for various defect centers.[40,41] The two spin-conserving transitions can be individually addressed in an external magnetic field, owing to the different effective *g*-factors in these orbital states. In twisted bilayer PbS, $g_{xy}=g_{yx}=2.37$ and $g_{xy}=g_{yx}=1.63$ for the states with $l_z=\pm 1$, respectively. $g_{xy}=g_{yx}=2$ in the $p_z$-state. (Details seen in supplementary Note 5). After the electron injection, the system is initialized in its negative charge state and spin-half spin state. The electron injection is optically detected by driving $\alpha$ and $\beta$ transitions (optically active negative charge state and inactive neutral charge state). The successful charge state preparation is ensured by the post-selection of optically active qubits. High population polarization to $|\downarrow\rangle$ spin state is achieved by electron spin hyperpolarization. By driving the $\beta$ optical pumping, the population in $|\uparrow\rangle$ is depleted owing to the optically forbidden spin-flip relaxation to the ground state. The achievable operation temperature is limited by the ratio of the optical lifetime and the temperature-dependent $T_1$ spin relaxation time.

Optical spin state readout is on one hand possible by monitoring the photoluminescence signal contrast of the $\beta$ cycling transition. However, this continuous-wave readout signal can deteriorate due to the spin-flip transition. Single-shot optical readout can improve the detection fidelity, as demonstrated for the NV center in diamond.[42] It relies on the optical detection of the spin-charge conversion of the system. The latter is induced by resonant optical ionization of the excited $|\uparrow\rangle$ spin state. As only the negative charge state is optically active, the spin-charge conversion can be detected by losing the $\alpha$ and $\beta$ photoluminescence signal. The protocol is analogous in the case of fully occupied dispersionless-band systems, such as twisted bilayer $Bi_2Se_3$ and twisted $In_2Se_3$.

Possible sources of decoherence include nuclear spin, thermal phonon noise, black-body radiation, and spontaneous emission. Nuclear spin, in particular, significantly affects the spin decoherence time $T_2$ of a qubit. Among all Pb and S isotopes, only $^{207}$Pb has a nuclear spin of 1/2 (See Supplemental Table 2). Fortunately, decoherence from $^{208}$Pb can be significantly diminished by utilizing a nuclear-spin-free isotope. On the other hand, addressable hyperfine transitions can be utilized to transfer and store the electron spin qubit to a nearby nuclear spin qubit. The latter is expected to serve as a long-living memory for the quantum computer due to the much smaller coupling strength of the nuclear spin to stray magnetic fields.

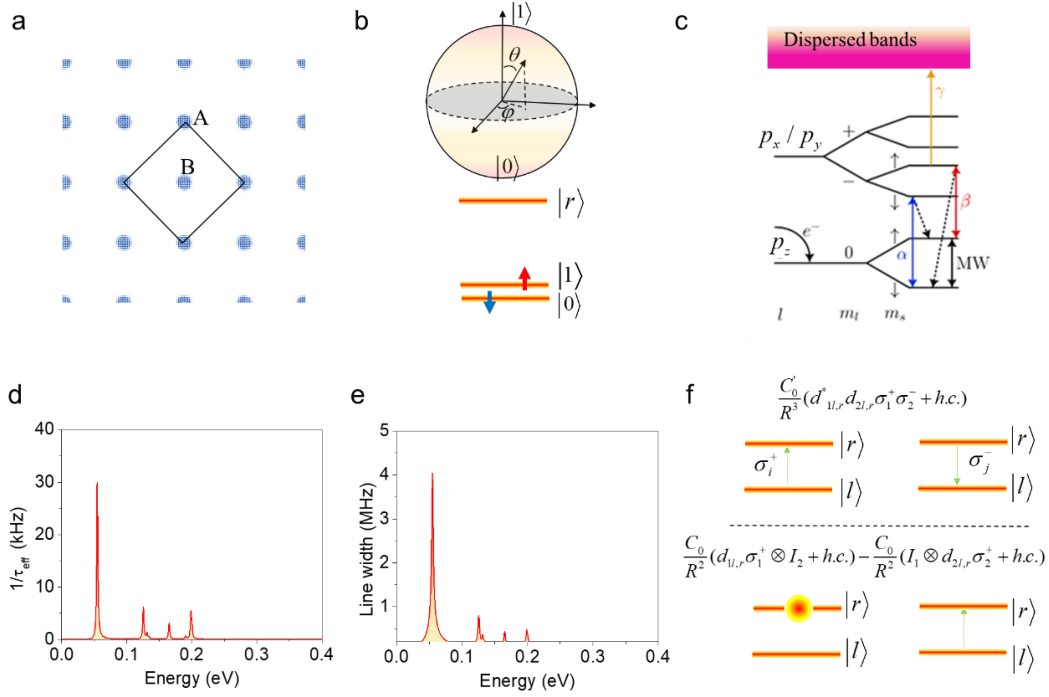

**Figure 3. Qubit realization. a** DFT-calculated periodic local states. Diamond frame implies the unit cell. **b** Encoding subspace and Bloch sphere representation. **c** Resonant optical pumping between the Zeeman levels of the electric dipole allowed transition are labeled $\alpha$ and $\beta$. Spin selection rule forbidden transitions are shown with dashed arrows. Microwave spin manipulation in the ground state is labeled by MW. $\gamma$ transition to the dispersed bands optically ionizes the system to its neutral state where it loses its optical activity. **d** Calculated lifetime of excited states. **e** Calculated linewidth of excited states. Temperature is set at 1 K. **f** Illustration of two-qubit interaction.

At high temperatures, decoherence primarily arises from thermal phonon noise. Detailed derivations of phonon-induced line broadening are outlined in Supplementary Note 6. The temperature only shifts the energy of flat bands but does not affect the band dispersion or the relative energy separation as shown in supplementary Fig. S6. Furthermore, the energy shift can be suppressed by decreasing the temperature. Similar to alkali-atom systems, decoherence can also result from black-body radiation and spontaneous emission. In the case of zero phonons, the calculated lifetimes and linewidths are depicted in Fig. 3d&e, with detailed calculations provided in the methods section. At 1 K, the calculated lifetime of the excited state is approximately 100 μs, much larger than that of alkali atoms. The linewidth of an excited state is around 10 MHz, significantly smaller than the spin splitting of the computational subspace. As temperature decreases, both linewidth and lifetime increase. The decoherence mechanisms here are analogous to those in qubits based on solid-state materials, which have been shown to function effectively at 1 K or even above.[41,43]

## Two-qubit interaction

To perform a two-qubit operation, interaction and entanglement between two qubits are essential. We achieve controllable interaction using an excited state similar to the case of alkali atoms. The three-level system is depicted in Fig. 3b, and the interaction is illustrated in Fig. 3f. After the Coulomb interaction was truncated at the third-order terms (detail seen in Supplementary Note 7), Hamilton of qubits is as follows

$$H = \frac{\hbar\Omega}{2}\sum_i \tau_x^i + \frac{\Delta}{2}\sum_i \tau_z^i + \sum_{i<j} \frac{C_0'}{R^3}(d^*_{il,r} d_{jl,r} \tau_i^+ \tau_j^-) + \frac{C_0}{R^2}(d_{il,r}\tau_i^+ \otimes I_j - I_i \otimes d_{jl,r}\tau_j^+) + h.c., \quad (3)$$

where the Pauli matrix $\tau$ acts on the basis set of a $p_z$-state and an excited state, and $I$ is identity matrix. $d$ is dipole matrix. The spin is temporally neglected. $\Omega$ is the driving frequency of a laser, and $\Delta$ is the detuning. The first and second terms are operations on a single qubit. The third term is the charge-dipole interaction, while the last term is the dipole-dipole interaction. The third term decays with distance as $R^{-2}$, and the fourth term decays with distance as $R^{-3}$. In twisted bilayer PbS, $C_0 = 1.44 eVnm^2$ and $C_0' = 1 eVnm^3$. The interaction strength is significantly larger than the dipole-dipole interactions in alkali-atom systems.

**Advantages of moiré qubits over other typical qubits**

Compared to other systems, moiré systems offer several advantages, as summarized in Extended Data Table 2. The quantum dots within moiré systems form a self-organized two-dimensional matrix, which inherently facilitates scalability for qubits. The quantum dots are uniform, and the inter-dot spacing is tunable through the twisting, allowing for adjustable interactions. Moreover, electrostatic gates can selectively address any individual quantum dot. The wavefunctions associated with flat bands exhibit high symmetry, and optical transitions adhere to specific selection rules that minimize unexpected transitions. Resonantly addressable optical transitions are crucial for qubit operation at elevated temperatures, enabling non-equilibrium population states and high-fidelity readouts. Additionally, spin-selective optical activity in a scalable qubit system supports long-distance quantum network applications. Our findings demonstrate that a spin-photon interface is feasible within moiré qubits for quantum repeater applications. This interface, in principle, facilitates connections to photonic qubit systems, paving the way for hybrid qubit platforms.

**Methods**

Density functional theory calculation

DFT is a numerical method for calculating the electronic structure of materials based on the first principles of quantum mechanics. Many DFT code packages are well developed and tested, achieving very high precision compared with experimental results.[44] In our calculations, we employ the PWmat code package, a well-established tool in the materials research community. The calculations utilize a single-zeta atomic basis set and FHI pseudopotentials. The PBE exchange-correlation functional is applied to handle the electron-electron exchange interaction within the mean-field theory framework. Additionally, DFT-D2 is employed to account for interlayer molecular interactions. Due to the large lattice length, 1×1×1 k-mesh is applied when the twisted angle is below 4.58°. Density-mesh is cut off at 80 Hartree. The spin-orbital coupling is neglected. The atomic positions in the twisted structures are relaxed until the force on each atom is smaller than 0.01 eV/Å. We have tested the GGA+U method with nonzero U for the structure with a twist angle above 4.58°, observing no significant differences.

Calculation of moiré potential

Under a small twist angle, $V(\vec{r})$ is usually approximated by the maximal fluctuation of the band edge in the real space when two layers are shifted.[29] This is the so-called moiré potential, and its maximum value is approximated as quantum-well depth.

Calculation of lifetime

The effective lifetime $\tau_{eff}$ is determined as $1/\tau_{eff} = 1/\tau_0 + 1/\tau_{BBR} + 1/\tau_{ep}$, where $\frac{1}{\tau_{BBR}} = \sum_{n'l'} \frac{A_{nl \to n'l'}}{e^{\hbar \omega_{nn'}/(k_B T)} - 1}$, $A_{nl \to n'l'}$ is the dipole matrix element for transitions between two states $nl$ and $n'l'$. $\frac{1}{\tau_0} = \sum_{n'l'} A_{nl \to n'l'}$ with the summation over all dipole-coupled states with energy $E_{n'l'} < E_{nl}$. All dipole matrix elements are extracted from the DFT calculations.

At high temperatures, decoherence primarily arises from thermal phonon noise. To mitigate the error of DFT, we simulate the molecular dynamics of twisted bilayer layer PbS at 100 K. The average phonon energy is about 0.8 meV/atom, translating to an electron energy of approximately 10.5 meV due to the weak electron-phonon coupling (shown in Supplementary Fig. 6)

Discussion on experimental synthesis and protection

Although moiré superlattices are referred to as twisted bilayer materials, their fabrication in experiments does not involve twisting. Instead, moiré superlattices are prepared by stacking two layers, one on top of the other.[45] Alternatively, moiré superlattices can be synthesized using chemical vapor deposition.[30,46] The fabrication of twisted bilayer materials is thus easy to control, even for non-vdW twisted layers. Twisted bilayer materials are atomically planar and lack dangling bonds. Qubits based on moiré systems are not susceptible to interference from external molecules, ions, and nanoparticles in the environment. Moreover, any influence from such entities can be mitigated by protective coatings, such as layers of hexagonal boron nitride. We believe our study will pave a new pathway for creating solid-state qubits and multi-qubit platforms for quantum information science.

**Data availability**

The authors declare that the main data supporting the findings of this study are available within the paper and its supplementary files. Other relevant data are available from the corresponding authors upon reasonable request.

**Author Contributions**

Corresponding author

Correspondence[*] to Zhigang Song (szg@pku.edu.cn) or Prineha Narang (prineha@ucla.edu).

**Acknowledgements**

This work is supported by the U.S. Department of Energy, Office of Science, Basic Energy Sciences (BES), Materials Sciences and Engineering Division under FWP ERKCK47 'Understanding and Controlling Entangled and Correlated Quantum States in Confined Solid-state Systems Created via Atomic Scale Manipulation'.

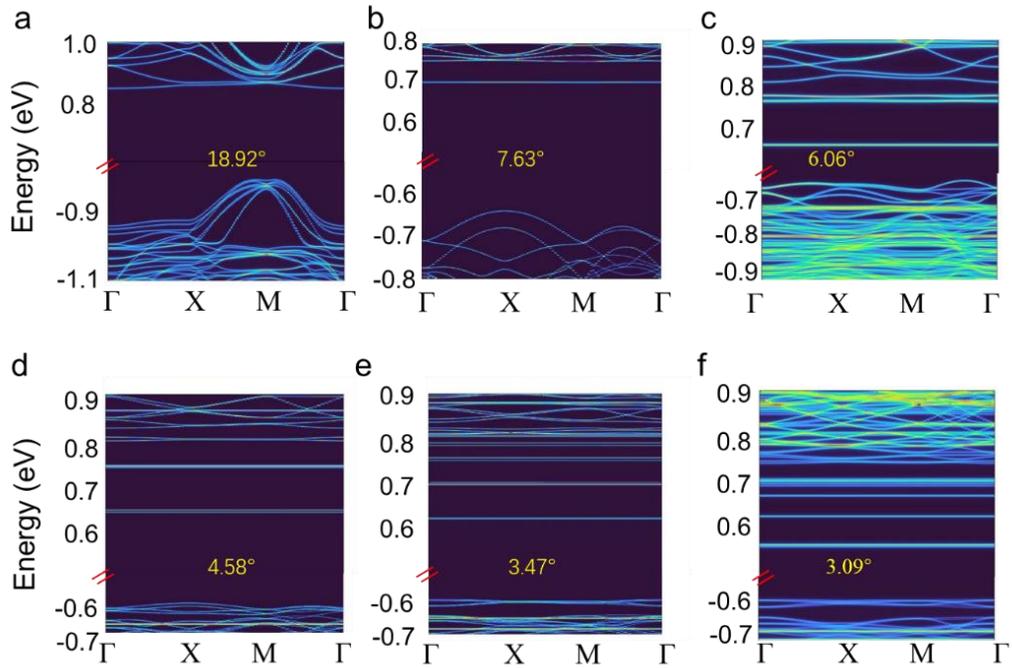

**Extended Data Figure 1. Band structure of bilayer PbS with different twist angles.**

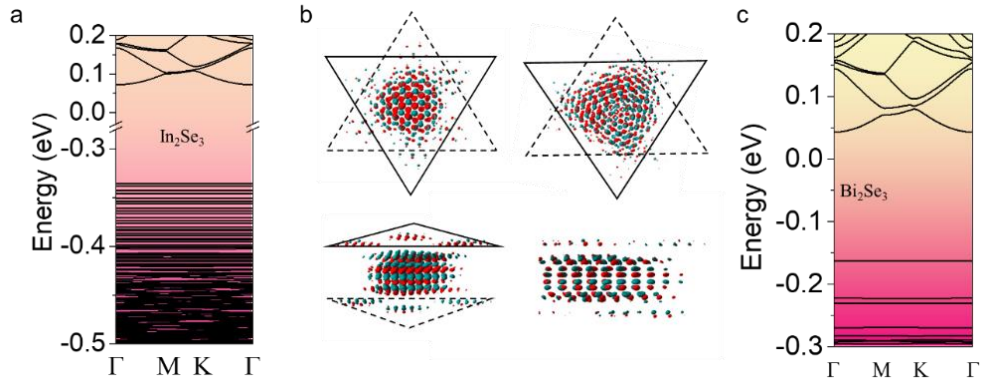

**Extended Data Figure 2. Electronic structure with flat bands at occupied states. a** Band structure of antiferroelectric twisted bilayer $In_2Se_3$ with a twist angle of 2.00°. **b** Calculated wavefunction of the first dispersionless band below the Fermi level in twisted bilayer $In_2Se_3$ (left column) and $Bi_2Se_3$ (right column). **c** Band structure of twisted bilayer $Bi_2Se_3$ with a twist angle of 2.00°.

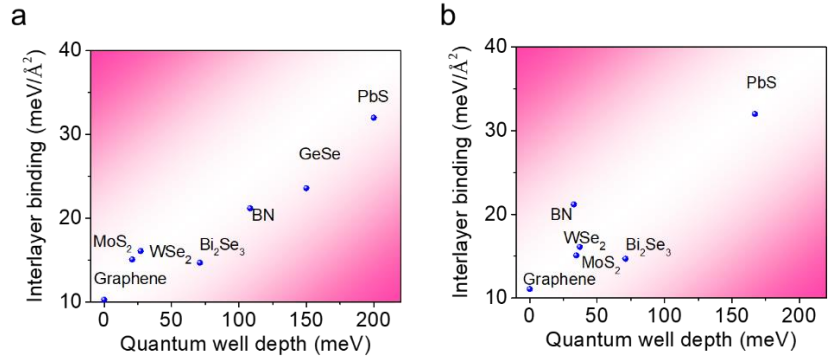

**Extended Data Figure 3. Quantum well depth in twisted materials versus interlayer binding energy.** Effective potential depth for electrons at conduction band minimum **a** and valence band maximum **b** among different materials.

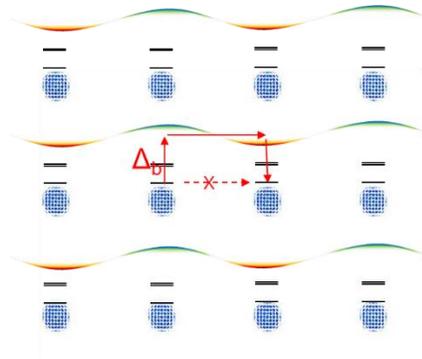

**Extended Data Figure 4. Illustration of tunneling paths and barriers for electron tunneling between different quantum dots.** Solid red arrows indicate possible electron tunneling paths, while dashed arrows represent forbidden paths. Black lines denote the flat bands of local states within a quantum dot. $\Delta_b$ denotes the tunneling barrier.

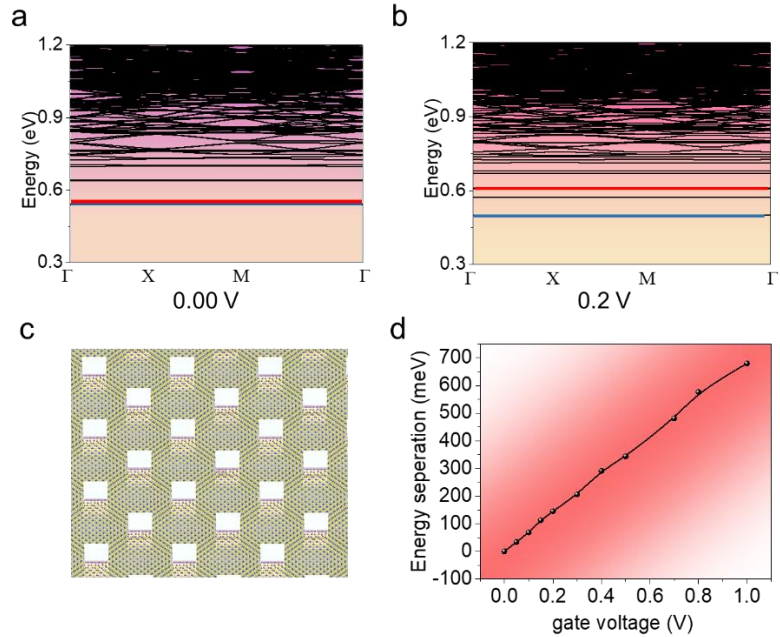

**Extended Data Figure 5. DFT-calculated control efficiency of electrostatic gates. a** Band structure of twisted bilayer PbS with a zero gate voltage. **b** Band structure with a gate voltage of 0.15 V. The two $p_z$-states between the two neighbor quantum dots are highlighted in red and blue, respectively. **c** Illustration of fabricated local electrostatic gates. **d** Energy separation between two $p_z$-states of neighboring quantum dots as a function of the applied electrostatic gate voltages. The twist angle is 2.66°. The dielectric layer is assumed to be an $Al_2O_3$ layer with a thickness of 1.0 nm.

| $D_4$ | E | $2C_4(z)$ | $C_2(z)$ | $C'2$ | $C''2$ | Linear functions | Quadratic functions | Cubic functions |
|---|---|---|---|---|---|---|---|---|
| $A_1$ | +1 | +1 | +1 | +1 | +1 | - | $x^2+y^2, z^2$ | - |
| $A_2$ | +1 | +1 | +1 | -1 | -1 | z | - | $z_3, z(x^2+y^2)$ |
| $B_1$ | +1 | -1 | +1 | +1 | -1 | - | $x^2-y^2$ | xyz |
| $B_2$ | +1 | -1 | +1 | -1 | +1 | - | xy | $z(x^2-y^2)$ |
| E | +2 | 0 | -2 | 0 | 0 | (x,y) | (xz,yz) | $(xz^2,yz^2)(xy^2,x^2y)(x^3,y^3)$ |

**Extended Data Table 1    Character table of the $D_4$ point group**

| | Natural atoms | | Artificial atoms | | | | |
|---|---|---|---|---|---|---|---|
| | Neutral atoms | Trapped ions | Molecule | Superconductor circuits | Semiconductor quantum dot | Color center | Moiré qubit |
| Energy gap | $10^5$ GHz | $10^5$ GHz | $10^5$ GHz | 1-10 GHz | $10^4$ GHz | $10^5$ GHz | $10^4$ GHz |
| Dimension | ~2Å | ~2Å | ~2nm | ~μm | ~nm | ~μm | ~2 nm |
| Distance | <1μm | ~5μm | - | ~μm | 10-100nm | ~10nm | 10-10000 nm |
| Operating temperature | nK-μK | μK-mK | ~K | ~mK | ~mK | 300K | 1K |
| Qubit interaction | Collision exchange | Coulomb | Coulomb Exchange, dipolar | Capacitive induce | Coulomb Exchange, dipolar | Coulomb Exchange, dipolar | Coulomb Exchange, dipolar |
| Ground Life time | ~h | ~h | ~h | ~μs | ~h | ~ms | ~h |
| Excited Life time | ~ms | ~s | ~μs | ~μs | - | ~μs | ~100μs |
| Linewidth | ~10MHz | ~10MHz | ~10GHz | ~100 MHz | - | - | ~10MHz |
| Scalability | good | good | excellent | good | poor | poor | Excellent |
| Identical qubits | excellent | excellent | poor | poor | poor | poor | Excellent |

**Extended Data Table 2.   Comparison of physical parameters between different qubit systems** [47,48]